\documentclass[%
 reprint,
twocolumns
 amsmath,amssymb,
 aps,
pra,
]{revtex4-2}
\newcommand{\be}{\begin{equation}}
\newcommand{\ee}{  \end{equation}}
\newcommand{\ba}{\begin{eqnarray}}
\newcommand{\ea}{  \end{eqnarray}}

\usepackage{graphicx}
\usepackage{float}
\usepackage[dvipsnames]{xcolor}
\usepackage{bbm}
\usepackage{amsmath, physics}
\usepackage{bbold}
\usepackage{mathtools}
\usepackage{dcolumn}

\usepackage{subfig}

\usepackage{bm}
\usepackage{mathtools}
\usepackage{multirow}
\usepackage[colorlinks=true,linkcolor=blue,citecolor=blue,urlcolor=blue]{hyperref}

\usepackage{cleveref}

\RequirePackage[normalem]{ulem}
\usepackage{color}

\begin{document}

\title{Tunneling in double-well potentials within Nelson's stochastic mechanics: Application to ammonia inversion}

\author{Danilo F. Schafaschek}
\affiliation{Departamento de F\'isica, Universidade Federal do Paran\'a, Curitiba, 81531-980, PR, Brazil}
\author{Giovani L. Vasconcelos}
\affiliation{Departamento de F\'isica, Universidade Federal do Paran\'a, Curitiba, 81531-980, PR, Brazil}
\author{Ant\^onio M. S. Mac\^edo}
\affiliation{Laborat\'orio de F\'isica Te\'orica e Computacional, Departamento de F\'isica, Universidade Federal de Pernambuco, Recife, 50670-901, PE, Brazil}


\date{\today}

\begin{abstract} 
Nelson's stochastic mechanics formulates quantum dynamics as a real-time conservative diffusion process in which a particle undergoes Brownian-like motion with a fluctuation amplitude fixed by Planck’s constant. While being mathematically equivalent to the Schrödinger formulation, this approach provides an alternative dynamical framework that enables the study of time-resolved quantities that are not straightforwardly defined within the standard operator-based approach.  In the present work, Nelson's stochastic mechanics is employed to investigate tunneling-time statistics for bound states in double-well potentials. Using first-passage time theory within this framework, both the mean tunneling time, $\bar{\tau}$, and the full probability distribution, $p({\tau})$, are computed. The theoretical predictions are validated through extensive numerical simulations of stochastic trajectories for two representative potentials. For the square double-well potential, analytical expressions for $\bar{\tau}$ are derived and are shown to be in excellent agreement with simulations. In the high-barrier limit, the results reveal a direct relation between the stochastic-mechanical and quantum-mechanical tunneling times, expressed as $\tau_{\mathrm{QM}} = (\pi/2)\bar{\tau}$, where $\tau_{\mathrm{QM}}$ corresponds to half the oscillation period of the probability of finding the particle in either well. This relation is further confirmed for generic double-well systems through a WKB analysis. As a concrete application, the inversion dynamics of the ammonia molecule is analyzed, yielding an inversion frequency of approximately $24$~GHz, in close agreement with experimental observations. These results highlight the potential of  stochastic mechanics as a conceptually coherent and quantitatively consistent framework for analyzing tunneling phenomena in quantum systems. \end{abstract}


\maketitle


\section{Introduction}

Because time is not represented by an operator in quantum mechanics---but rather appears only as a parameter labeling the evolution of quantum states---the problem of defining, let alone measuring, the traversal time between two points for a quantum particle remains an elusive and, to some extent, controversial issue 
\cite{HaugeStovnengRMP,LandauerMartinRMP1994, Hilgevoord2005,WinfulPhysRep2006,MugaBookVol1}.
The question becomes particularly subtle in the context of quantum tunneling, where a particle traverses a potential barrier higher than its own energy---a process forbidden in classical dynamics. This raises fundamental questions, such as how long it takes for the particle to tunnel through the barrier or, conversely, how much time it spends within the classically forbidden region. 

In order to address these and related questions, a variety of operational and theoretical proposals for ‘traversal time’ have been advanced. The {dwell time} defines the average time spent by the particle in a specified spatial region, irrespective of whether it is ultimately transmitted or reflected~\cite{HaugeStovnengRMP}. {The phase delay time} 
extracts a time scale from the energy derivative of the scattering phase (or of the $S$-matrix), and is often applied to the transmitted component in one-dimensional tunneling~\cite{Wigner1955,Smith1960,DeMouraAlbuquerque_SSC_1990}.
The {B\"uttiker--Landauer (BL) time} probes traversal by monitoring the system's sensitivity to a slow modulation of the barrier, yielding an operational time linked to adiabatic response~\cite{BuettikerLandauerPRL,BuettikerPRB}. {Larmor-clock times} introduce a weak, localized magnetic field in the barrier and infer the time from spin precession~\cite{Buettiker1990,LandauerMartin1992}.
Bohmian (trajectory-based) analyses likewise define traversal and residence times from guidance trajectories, offering a complementary, causal picture that can be directly compared with operational proposals~\cite{LeavensBohmian,OriolsMompart}.
More recently, alternative formulations have sought to promote time to the status of a genuine quantum observable, such as the space-time-symmetric extensions of quantum mechanics that introduce a self-adjoint time operator and a dual, space-conditional Schrödinger equation within the same Hilbert-space framework~\cite{DiasParisioPRA2017,AraujoParisioPRA2024,FloresTOA2024,deLaraBeims2024_TravelingTime}.
Despite this profusion of proposals, no single notion of traversal (or tunneling) time has gained universal acceptance: each highlights a particular operational or theoretical aspect of the process, yet each also entails limitations that preclude it from serving as a context-independent notion of “time”~\cite{HaugeStovnengRMP,LandauerMartinRMP1994,OlkhovskyRecamiPR2004,WinfulPhysRep2006,MugaBookVol1,TimeInQMVol2}.

Stochastic mechanics (SM) is a formalism introduced by Nelson in 1966 that describes quantum systems as \textit{conservative  diffusion processes}~\cite{Nelson1966}. 
In Nelson's framework, the dynamics of a quantum particle is governed by a stochastic differential equation whose deterministic drift term accounts for the interaction potential, while the diffusion 
coefficient (related to the strength of the underlying Brownian motion) is proportional to Planck’s constant. 
By imposing a stochastic analogue of Newton’s second law, Nelson demonstrated that stochastic mechanics is mathematically equivalent to the Schrödinger equation. 
Subsequent developments
~\cite{YasueJFA1981,YasueJMP1981, GuerraMoratoPRD1983,
PavonJMP1995,YangJMP2021} clarified the Lagrangian and Hamiltonian   variational underpinnings of stochastic mechanics and its unitary correspondence~\cite{Pavon2000}  with conventional quantum mechanics. For both earlier and more recent reviews of stochastic mechanics,  see, e.g., Refs.~\cite{Guerra1981, ZambriniIJTP1985, Nelson1985, BlanchardCombeZheng1987, BeyerPaulUniverse2021,Kuipers2023}.
Nelson’s formalism may also be viewed, in hindsight, as a dynamical counterpart of  Schrödinger bridge problem introduced in 1931~\cite{Schrodinger1931}. In that framework, one selects---via a relative-entropy minimization---the most probable diffusion process interpolating between prescribed endpoint distributions. This problem is now understood as a precursor to modern mass-transport theory~\cite{Christian2014,Mikami2021}; see also Ref.~\cite{Chetrite2021} for an English translation of Schrödinger’s original paper and a detailed discussion of its connections to large-deviation theory and stochastic optimal control.


It should be noted that the formal equivalence between SM and the Schrödinger equation has been critically examined in the literature. In particular, Wallstrom~\cite{Wallstrom1989,Wallstrom1994} observed that formulations expressed solely in terms of the hydrodynamic variables---namely, the probability density and the phase (see below)---may admit solutions that do not automatically satisfy the single-valuedness condition of the Schrödinger wavefunction, thereby indicating that an additional quantization condition on the phase may be required. Subsequent analyses have argued that appropriate regularity, topological, or variational considerations remove the spurious solutions; see, e.g., Ref.~\cite{Kuipers2023}. In the present work we restrict our attention to states corresponding to standard Schrödinger solutions, and thus operate entirely within the conventional quantum sector of the theory.

A related concern, already emphasized by Nelson \cite{Nelson1985}, pertains to the apparent mismatch between multi-time correlations of the underlying diffusion process and the operator-ordered correlation functions of textbook quantum theory. This issue has been analyzed in detail by Blanchard {\it et al.}~\cite{Blanchard1986PRD}, who showed that a proper treatment of measurement-induced state updating within the stochastic framework restores agreement with standard quantum predictions; more recent discussions have further clarified this point within both operational and modified dynamical formulations of SM~\cite{Derakhshani2024,Kuipers2023}.
Nelson also emphasized that, for entangled systems, the stochastic drift in configuration space exhibits explicit nonlocal dependence~\cite{Nelson1986FieldTheory}. From a contemporary perspective, however, this feature is not generally regarded as a defect but rather as a structural property shared by realist trajectory-based reconstructions of quantum mechanics that successfully reproduce entanglement phenomena. In this sense, the nonlocal character of SM may be viewed as reflecting its compatibility with the implications of Bell-type results rather than signaling an internal inconsistency of the theory; see also recent discussions in Refs.~\cite{Derakhshani2024,Kuipers2023}.

Because the SM formalism is expressed in terms of real-space stochastic trajectories, it provides a natural framework for investigating dynamical quantities that are difficult to define or access within the standard quantum-mechanical formulation. 
One such quantity is precisely the \textit{traversal time} between two spatial points. 
Indeed, by employing the concept of \textit{first-passage times} from the theory of diffusion processes~\cite{Gardiner2009}, one can evaluate not only the mean traversal time but also its full probability distribution---a quantity rarely addressed by other tunneling-time proposals discussed above. 
These aspects make stochastic mechanics a powerful tool for studying tunneling phenomena, as demonstrated in previous works on tunneling through square barriers~\cite{ChenWang_PLA_1990,Hara2003,KudoNittaPLA2013} and on bound-state tunneling in a quartic double-well potential~\cite{KoeppeJMP2018_QHE_1D}. 

In this paper, we employ Nelson's stochastic mechanics to investigate in detail the tunneling process in double-well potentials, where a particle passes through the barrier separating the two minima. We compute the mean tunneling time, denoted by $\bar{\tau}$, 
for each potential considered and analyze its dependence on relevant physical parameters. We show, in particular, that $\bar{\tau}$ has a clear physical interpretation, as it can be directly related to the corresponding—and spectroscopically measurable—``tunneling time'' derived from standard quantum mechanics within the two-level approximation. We further show that the empirical distributions of tunneling times obtained from large ensembles of numerical trajectories exhibit exponential tails that are in excellent agreement with the theoretical predictions derived from first–passage time theory.
Specifically, we first consider a square double-well potential, for which the mean tunneling time can be calculated analytically. Theoretical predictions are shown to be in excellent agreement with numerical simulations, thereby validating the analytical approach. We also analyze the high-barrier limit, establishing a direct relation between the tunneling time predicted by SM and the corresponding tunneling time obtained from quantum mechanics (see below). We then demonstrate, through a WKB calculation, that this relation holds for generic double-well systems.
Finally, we apply the SM framework to the inversion problem of the ammonia molecule, in which the nitrogen atom oscillates between two equivalent configurations on opposite sides of the plane defined by the three hydrogen atoms—effectively tunneling through the potential barrier separating the two states.

In quantum mechanics, tunneling in a double-well potential is typically treated within the two-state approximation. By constructing symmetric and antisymmetric superpositions of the ground and first excited states, one obtains states localized primarily in the left and right wells, respectively. If the system is initially prepared in one of these localized (nonstationary) states, the probability of finding the particle in either well oscillates periodically in time, with period~$T$ inversely proportional to the energy splitting~$\Delta E$. The oscillation half-period,
$\tau_{\mathrm{QM}} = {T}/{2}$,
is commonly referred to as the quantum-mechanical tunneling time. 
We demonstrate—first analytically for the square double-well potential and subsequently through a WKB analysis for generic double-well systems—that, in the high-barrier limit, the following relation between the SM mean tunneling time~$\bar{\tau}$ and the quantum-mechanical tunneling time~$\tau_{\mathrm{QM}}$ holds:
\begin{equation}
\tau_{\mathrm{QM}} = \frac{\pi}{2}\bar{\tau}.
\label{eq:1}
\end{equation}
This relation, previously observed numerically for the quartic double-well potential~\cite{BeyerPaulUniverse2021}, is here derived analytically and shown to hold generally, provided the barrier is sufficiently high.

Having established these theoretical results, we next address the inversion problem of the ammonia molecule as a concrete tunneling example. To this end, we compute the tunneling times for a particle moving in a double Rosen–Morse potential, a model commonly employed for the ammonia inversion process~\cite{RosenMorse,YangHan_IJMS_2021_Ammonia}.
Although this case cannot be solved analytically, the mean tunneling time can be obtained numerically by direct integration of the corresponding SM expression. From the SM mean tunneling time~$\bar{\tau}$ and relation~(\ref{eq:1}), we estimate the inversion frequency of the ammonia molecule. The resulting value—approximately $24$~GHz—is in excellent agreement with experimental observations, thus validating the SM-based theoretical approach to tunneling-time calculations. 

The remainder of this paper is organized as follows. 
Section~II provides a brief review of stochastic mechanics. It also 
discusses how the mean tunneling time and the full distribution of tunneling times can be obtained from the theory of first-passage times within the SM framework. 
Section~III applies the SM formalism to tunneling in a square double-well potential, where most calculations can be performed analytically, including the derivation of Eq.~(\ref{eq:1}). 
Similar calculations are then extended in Sec.~IV, using a WKB approach to treat generic double-well systems. 
Section~V presents a concrete application of the SM approach to the inversion problem of the ammonia molecule.
The implications of our findings are discussed in Sec.~VI. Section~VII summarizes our main conclusions.

\section{Stochastic Mechanics: Brief Review}

\label{sec:review}

Although introduced nearly six decades ago, Nelson’s formulation of stochastic mechanics (also known as \textit{stochastic quantization}) is perhaps not widely known among physicists. For this reason, we include here a brief, self-contained exposition of its basic principles.
Our treatment departs slightly from Nelson’s original presentation \cite{Nelson1966} 
in that we demonstrate how one of Nelson’s dynamical conditions [see Eq.~\eqref{eq:v}] can be derived directly from the requirement of time reversibility, whereas in Nelson’s original formulation it was introduced as an {\it ad hoc} assumption.

In  Nelson's stochastic mechanics approach, quantum states are described by the  following pair of forward and backward stochastic differential equations:
\begin{align}
d  \mathbf{x}&=  \mathbf v_+(\mathbf x,t) dt+ {\sigma}  d\mathbf{W}(t) ,
    \label{eq:dx1}\\
d  \mathbf{x}&=  \mathbf v_-(\mathbf x,t) dt+ {\sigma}  d\mathbf{W}_*(t) ,
    \label{eq:dx2}
\end{align} 
where $\mathbf v_\pm (\mathbf x,t)$ are the forward and backward drift velocities to be determined later and $\mathbf{W}(t)$ and $\mathbf{W}_*(t)$ are forward and backward Wiener processes, respectively, such that
$d \mathbf W(t)$ is independent of all $\mathbf x(s)$
with $s < t$, 
whereas $d \mathbf W_*(t)$ is independent of all $\mathbf x(s)$
with $s > t$.
In order to recover quantum mechanics from the stochastic dynamics above, it is necessary to determine specific conditions for the drift coefficients $\mathbf v_\pm(\mathbf x,t)$ and the noise amplitude $\sigma$, as discussed next.

\subsection{Hydrodynamical Equivalence to Quantum Mechanics}

To ensure time reversibility, both processes in \eqref{eq:dx1} and \eqref{eq:dx2} must yield the same  probability density $\rho(\mathbf x, t)$.
In other words, $\rho(\mathbf x, t)$  must satisfy both the  forward and  backward Fokker-Planck equations associated with the processes in (\ref{eq:dx1}) and (\ref{eq:dx2}), namely:
\begin{align}
    \frac{\partial \rho}{\partial t}+ \mathbf{\nabla}\cdot\left[\mathbf v_ \pm \rho\mp\frac{{\sigma}^2}{2}  \nabla  \rho\right] =0.
    \label{eq:FK1}
\end{align}
Adding these two equations  yields  the continuity equation
\begin{align}
    \frac{\partial \rho}{\partial t}+ \mathbf{\nabla}\cdot (\rho \mathbf v)=0,
  \label{eq:contN}
\end{align}
where
\begin{align}
\mathbf v = \frac{\mathbf v_+ + \mathbf v_-}{2} 
\label{eq:v}
\end{align}
is the {\it advective} or {\it current velocity}. Similarly, subtracting the two equations in \eqref{eq:FK1} gives
 the condition
\begin{align}
\mathbf u = 
{\sigma^2}\mathbf{\nabla}\ln\sqrt{\rho}.
\label{eq:u1}
\end{align}
where 
\begin{align}
\mathbf u = \frac{\mathbf v_+ - \mathbf v_-}{2} 
\label{eq:u}
\end{align}
is the so-called  {\it diffusive} or {\it osmotic velocity}. 
In terms of the  current and osmotic velocities, $\mathbf v$ and $\mathbf u$, the drift velocities $\mathbf v_\pm$ in Eqs.~\eqref{eq:dx1} and \eqref{eq:dx2} are given by
\begin{align}
\mathbf v_\pm =  \mathbf v \pm \mathbf u   .
\label{eq:vpm}
\end{align}
Using Eq.~(\ref{eq:u1}), the continuity equation (\ref{eq:contN}) can  be written explicitly in terms of the `velocity fields' $\mathbf v$ and $\mathbf u$: 
\begin{equation}
\frac{\partial \mathbf u}{\partial t}
   +\frac{\sigma^{2}}{2}\,\nabla\bigl(\nabla\!\cdot\mathbf v\bigr)
        + \nabla\!\bigl(\mathbf v\!\cdot\mathbf u\bigr)=0.
\label{eq:dudt3}
\end{equation}



The trajectories described by (\ref{eq:dx1}) and (\ref{eq:dx2}) are continuous but nowhere differentiable, hence the instantaneous velocity of the process is not defined pointwise. To circumvent this difficulty, Nelson defined mean forward and backward time derivatives:
\begin{align}
    D_+ f(\mathbf x(t),t)& =\lim_{\Delta t \to 0}\mathbb{E}\left[\frac{f(\mathbf x(t+\Delta t),t+\Delta t)-f(\mathbf x(t),t)}{ \Delta t}\right],\\
    D_- f(\mathbf x(t),t)& =\lim_{\Delta t \to 0}\mathbb{E}\left[\frac{f(\mathbf x(t),t)-f(\mathbf x(t-\Delta t),t-\Delta t)}{ \Delta t}\right],
\end{align}
where $\mathbb{E}$ denotes expectation value. Using Itô's calculus one can show that 
\begin{align}
    D_\pm f(\mathbf x(t),t)& =\left[\frac{\partial }{\partial t}+ 
    \mathbf v_\pm \cdot \mathbf \nabla \pm \frac{\sigma^2}{2}\nabla^2 \right]f(\mathbf{x}(t),t).
    \label{eq:Df}
\end{align}
In particular, applying \eqref{eq:Df} to (\ref{eq:dx1}) and (\ref{eq:dx2}) yields
\[D_\pm \mathbf x(t)=\mathbf v_\pm,\]  
so that 
\begin{align}
    \mathbf v &= \frac{1}{2}\left[D_+ + D_-\right]   \mathbf x(t) 
\end{align}
and 
\begin{align}
    \mathbf u &= \frac{1}{2}\left[D_+ - D_-\right]   \mathbf x(t) ,
\end{align}
which give a more physical interpretation to the quantities $\mathbf v$ and $\mathbf u$ as being indeed two types of mean velocities. 

Similarly, one can define two mean accelerations, as follows. The mean {\it advective} acceleration is defined as
\begin{align}
    \mathbf a_{\rm adv} 
    &= \frac{1}{2}\left[D_+ \mathbf v_-+ D_-\mathbf v_+\right];
\end{align}
whereas a `diffusive 
acceleration'  can be defined by
\begin{align}
    \mathbf a_{\rm diff} 
    &= \frac{1}{2}\left[D_+ \mathbf v_+- D_-\mathbf v_-\right].
\end{align}
Using (\ref{eq:Df}) one then finds that
\begin{align}
    \mathbf a_{\rm adv} &= 
    \frac{\partial \mathbf v }{\partial t}+\mathbf v\cdot \mathbf \nabla \mathbf v-\mathbf u\cdot \mathbf \nabla \mathbf u-\frac{\sigma^2}{2}\nabla^2\mathbf u,
     \label{eq:ad}
\end{align}
and
\begin{align}
    \mathbf a_{\rm diff} &= 
     \frac{\partial \mathbf u }{\partial t}+ \mathbf v\cdot \nabla\mathbf{u} + \mathbf u\cdot \nabla\mathbf{v}+\frac{\sigma^2}{2}\nabla^2\mathbf v,
     \label{eq:ao}
\end{align} 

Let us now consider the behavior of the kinematic quantities defined above, namely $\mathbf v$, $\mathbf u$, $\mathbf a_{\rm adv}$, and $\mathbf a_{\rm diff}$, under time reversal. Since $\rho(\mathbf x, t)$ is invariant under time reversal, it follows from (\ref{eq:contN}) and (\ref{eq:u1})  that $\mathbf v  \to - \mathbf v $ and $\mathbf u  \to  \mathbf u $ as $t\to -t$, thus
indicating that  $\mathbf v$ is a true velocity field, while $\mathbf u$ is a {pseudo velocity}.
Similarly, for the accelerations 
one has 
$\mathbf a_{\rm adv} \to  \mathbf a_{\rm adv}$, as expected for a true acceleration; whereas   $\mathbf a_{\rm diff} \to - \mathbf a_{\rm diff}$, showing that this quantity cannot represent a physically acceptable acceleration.
Thus, to ensure  a time-reversible dynamics we require that 
\begin{equation}
\mathbf a_{\rm diff}=0,
\label{eq:a0}
\end{equation} which yields  our first dynamical equation, namely
\begin{align}
    \frac{\partial \mathbf u }{\partial t}+ \mathbf v\cdot \nabla\mathbf{u} + \mathbf u\cdot \nabla\mathbf{v}+\frac{\sigma^2}{2}\nabla^2\mathbf v =0 .
     \label{eq:dudt}
\end{align} 
In order for both Eqs.~\eqref{eq:dudt3} and \eqref{eq:dudt} to hold, one can easily show that we must have
$\boldsymbol{\nabla}\times\mathbf{v}=0$. 
Hence we can write
\begin{align}
\mathbf{v} = \sigma^{2}\nabla S,
\label{eq:vS}
\end{align}
for some function $S(\mathbf{x},t)$. 
A related discussion is given in Ref.~\cite{delaPenaCetto2014}, where it is shown that postulating that $\mathbf{v}$ is irrotational leads to Eq.~\eqref{eq:a0}.
In contrast, here we have argued on physical grounds (time-reversal symmetry) that Eq.~\eqref{eq:a0} must hold, which in turn implies that $\mathbf{v}$ is irrotational.

In Nelson’s original derivation, Eq.~(\ref{eq:vS}) was postulated as an assumption.
This condition can also be justified within variational formulations of stochastic mechanics~\cite{GuerraMoratoPRD1983,PavonJMP1995}.
Now, using  that  $\mathbf v$  and $\mathbf u$  are both irrotational, Eq.~(\ref{eq:dudt}) can be written as
\begin{align}
    \frac{\partial \mathbf u }{\partial t}+ \nabla (\mathbf{u} \cdot \mathbf{v})+\frac{\sigma^2}{2}\nabla^2\mathbf v =0 .
     \label{eq:dudt1}
\end{align} 

Time reversibility   also has  an important implication for the possible allowed dynamics.  
To see this, let us take the curl of (\ref{eq:ad}). Using that both $\mathbf v$  and $\mathbf u$  are curl-free, one then immediately finds that $\mathbf \nabla \times \mathbf a_{\rm adv}=0$, which implies that
\begin{align}
 \mathbf a_{\rm adv} &= \mathbf \nabla \varphi(\mathbf x),
\label{eq:phi}
\end{align}
where $\varphi(\mathbf x)$ is an arbitrary function to be determined. Using (\ref{eq:phi}) in (\ref{eq:ad}), we then obtain  our second dynamical equation:
\begin{align}  
    \frac{\partial \mathbf v }{\partial t}+\frac{1}{2} \mathbf \nabla \mathbf v^2-\frac{1}{2} \mathbf \nabla \mathbf u^2-\frac{\sigma^2}{2}\nabla^2\mathbf u
   =\mathbf \nabla \varphi.
\label{eq:dvdt}
\end{align}

The remarkable discovery by Nelson \cite{Nelson1966} was to note that by considering the function \begin{equation}
\Psi(\mathbf x,t)=\sqrt{\rho(\mathbf x,t)}e^{iS(\mathbf x,t)},
\label{eq:psi}
\end{equation}
and setting 
\begin{align}
\sigma^2 = \frac{\hbar}{m}
\label{eq:sigma}
\end{align}
and
\begin{align}
 \varphi (\mathbf x) &= -\frac{V(\mathbf x)}{m},\label{eq:V}
\end{align}
where $V(\mathbf x)$ is the particle's potential energy function, the time-reversible stochastic dynamics described above   recovers the Schr\"odinger equation:
\begin{align}
    i\hbar \frac{\partial \Psi(\textbf{x},t)}{\partial t}= -\frac{\hbar^2}{2m}\nabla^2 \Psi(\textbf{x},t)+ V(\textbf{x})\Psi(\textbf{x},t).
    \label{eq:scho}
\end{align}
More precisely, 
with the identifications (\ref{eq:sigma}) and (\ref{eq:V}), 
Eqs.~(\ref{eq:dudt1}) and (\ref{eq:dvdt}) correspond precisely to the two flow equations  
obtained by  Madelung \cite{Madelung1927}  upon using
the representation (\ref{eq:psi}) in Schr\"odinger equation. 
In particular, with the identifications (\ref{eq:sigma}) and (\ref{eq:psi}) the flow velocities $\mathbf{v}$ and $\mathbf{u}$ read
 \begin{align}
\mathbf{v}(\mathbf{x}, t) &= \frac{\hbar}{m}\nabla S(\mathbf{x}, t) = \frac{\hbar}{m}{\rm Im}[\nabla\ln \Psi], \label{eq:vIm}\\
\mathbf{u}(\mathbf{x}, t) &= \frac{\hbar}{2m} \nabla \ln \rho(\mathbf{x}, t) = \frac{\hbar}{m}{\rm Re}[\nabla\ln \Psi]. \label{eq:uRe}
\end{align}

In summary, Nelson’s stochastic mechanics can be formulated in terms of two Newton-like dynamical laws that respect time-reversal symmetry, namely
\begin{align}
 \mathbf a_{\rm diff} &= 0,   \label{eq:Newton1}\\
 \mathbf a_{\rm adv} &= -\frac{1}{m}\mathbf \nabla V(\mathbf x);   \label{eq:Newton2}
\end{align}
together with the following kinematic relations:
\begin{align}
d  \mathbf{x}&=  (\mathbf v+\mathbf u) dt+ \sqrt{\frac{\hbar}{m}}  d\mathbf{W}(t) ,
    \label{eq:dx12}\\
d  \mathbf{x}&=  (\mathbf v-\mathbf u) dt+ \sqrt{\frac{\hbar}{m}} d\mathbf{W}_*(t) ,
    \label{eq:dx22}
\end{align} 
which are likewise time-reversal symmetric.

It is also possible to establish  the unitary equivalence between Nelson's stochastic mechanics and standard quantum mechanics, which provides a rigorous mathematical foundation for the 
correspondence between the two formalisms (see Appendix). This shows  that they are not merely analogous but 
are in fact related by a fundamental unitary transformation. This deep connection suggests that 
the stochastic and quantum perspectives are simply two different ways of looking at the same 
underlying physical reality.

\subsection{Tunneling Times: Mean and Distribution}

\label{sec:dist}

For the sake of generality, the foregoing discussion considered motion in three-dimensional space. For the purposes of the present work, however, it suffices to restrict our attention to one-dimensional motion. Accordingly, we shall henceforth assume that the particle moves only along the $x$ axis.

Nelson's quantization  approach has the advantage that one can draw upon the theory of stochastic processes to investigate certain properties that are not easily accessible via  standard quantum mechanics. One such quantity is the mean first passage time for a particle  to exit a given interval. To be specific, let us consider the mean time that the particle takes to leave an interval $(a,b)$, starting from an initial point $x$. For our purposes here, let us assume that points $a$ and $b$ are reflecting and absorbing boundaries, respectively. Then, using standard formula from the theory of stochastic processes \cite{Gardiner2009,PaulBaschnagel1999}, the mean first passage time (MFPT) for the particle to cross the right boundary is given by
\begin{align}
   \bar\tau(x) &=\frac{2m}{\hbar}\int_x^b \frac{1}{\phi(y)}\left(\int_a^y \phi(z) dz\right) dy ,
   \label{eq:tau0}
\end{align}
where
\begin{align}
\phi(x) &=  \frac{2m}{\hbar}\exp\left(\int_a^x  v_+(y) dy\right),
\label{eq:phiU}
\end{align}
where $v_+(x)=v(x)+u(x)$; see Eq.~\eqref{eq:vpm}.
For the particular case of interest here, namely tunneling through a barrier in a double-well potential, the system is in a bound state and hence the probability current is zero, that is, $j=0$, which implies that the current velocity is zero, i.e., $v=0$, and so $v_+=u$, as follows from (\ref{eq:vpm}). Using (\ref{eq:u1}) into (\ref{eq:phiU}),  we obtain $\phi(x)=p(x)/p(a)$, where $p(x)=|\psi(x)|^2$, with $\psi(x)$ being the solution of the time-independent Schr\"odinger equation for the corresponding (bound) state of the system at hand. Then (\ref{eq:tau0}) becomes
\begin{align}
   \bar\tau(x) &=\frac{2m}{\hbar}\int_x^b \frac{1}{p(y)}\left(\int_a^y p(z) dz\right) dy  .
   \label{eq:tau_ab}
\end{align}
In applying Eq.~(\ref{eq:tau_ab}) to compute mean tunneling times, the interval $(a,b)$ and the starting point $x$ are chosen accordingly,
as will become clear in our discussion of specific double-well potentials below.

The theory of first passage times enables us to study not only the MFPT, as shown above, but also  the distribution of tunneling times, as discussed next.
Let $f(x,t)$ be the probability that  the particle will exit the interval $(a,b)$ at time $t$, after  starting in $t=0$ from  point $x$. Then one has \cite{Gardiner2009,PaulBaschnagel1999}
\begin{equation}
   f(x,t)=- \frac{\partial P_{\text{in}}(x,t)}{\partial t} ,
    \label{eq:dPdt}
\end{equation}
where $P_{\text{in}}(x,t)$ is the so-called survival probability, that is, the probability that the particle
has not yet exited the interval by time $t$:
\[
P_{\text{in}}(x,t) = {\rm Prob}\left( \tau > t \mid x(0) = x \right).
\]
Here $\tau$ is the first exit time from $(a,b)$. The survival probability satisfies a Sturm-Liouville problem,
\begin{equation}
\frac{\partial P_{\text{in}}}{\partial t} = \mathcal{L}_x P_{\text{in}}
\label{eq:Ps}
\end{equation}
where \( \mathcal{L}_x \) is the following (self-adjoint) operator
\[
\mathcal{L}_x = u(x)\frac{\partial}{\partial x} + \frac{\hbar}{ 2m}\frac{\partial^2}{\partial x^2},
\]
and $P_{\text{in}}(x,t)$ satisfies  initial and terminal conditions: \( P_{\text{in}}(x,0) = 1 \) and \( P_{\text{in}}(x,\infty) = 0 \). For the tunneling problem, it is convenient to choose the left and right endpoints as reflecting and absorbing boundaries, respectively, as we will discuss later. For this case, we  thus have the following boundary conditions: 
    $[\partial P_{\text{in}}(x,t)/\partial x]_{x=a}=0$ and $ P_{\text{in}}(b,t) = 0$.

From the method of separation of variables, one immediately concludes that 
$P_{\text{in}}(x,t)$ can be written as an eigenfunction expansion, where each eigenmode decays exponentially in time.
Hence, for large $t$, the survival probability decays as
\begin{equation}
P_{\text{in}}(x,t) 
\propto e^{-\lambda_1 t}, \quad \text{for} \quad t \to \infty,
\label{eq:longtime}
\end{equation}
where  {$\lambda_1$ is the smallest eigenvalue of ${\cal L}_x$.}
Now, using (\ref{eq:dPdt}) we can conclude that the first passage time density  has an exponential tail for large $\tau$:
\begin{align}
  f(x,\tau) 
  \propto 
  e^{-\lambda_1 \tau}, \quad \text{for} \quad \tau \gg 1.  
  \label{eq:exp}
\end{align}

We will see in the remainder of the paper that  both the mean tunneling time, given by Eq.~(\ref{eq:tau_ab}), and the exponential tail behavior of the tunneling-time distribution, as predicted by Eq.~(\ref{eq:exp}), are in excellent agreement with results  obtained from numerical simulations of  Nelson's stochastic  process for specific potentials. (An exponential decay in the tunneling-time distribution was observed numerically for both the single-barrier case \cite{Hara2003} and the quartic double-well potential \cite{Koeppe2017}, although apparently without recognition of the theoretical basis for this behavior, as established above.)
As already mentioned, our main interest here concerns tunneling processes
in double-well potentials that bear direct application, for instance, to the problem of inversion oscillation of the ammonia molecule. We begin  by considering the double square-well potential, for which analytical calculations can be carried out.

\section{Double Square Well}\label{sec: DSW}

\begin{figure}
    \centering
    \includegraphics[width=0.9\linewidth]{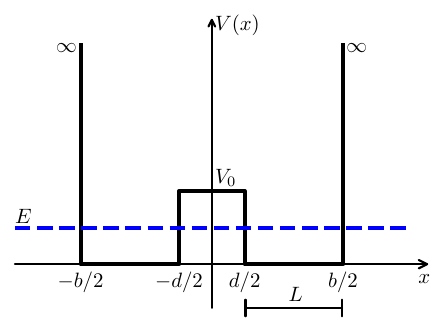}
    \caption{Double square well potential with arbitrary parameters. This potential can be built with two infinity walls at the positions $\{-\frac{b}{2}, \frac{b}{2}\}$ and with a barrier of thickness $d$ and height $V_0$ centered on the origin.}
    \label{fig: DSW-pot}
\end{figure}

We consider a double square well potential given by (see Fig.~\ref{fig: DSW-pot})
\begin{equation}
    V(x) = 
    \begin{cases}
    \infty, & |x| \geq b/2, \\
    V_0, & -d/2 < x < d/2, \\
    0, & \text{otherwise}.
    \end{cases}
    \label{eq:Pot}
\end{equation}
Solving the time-independent Schrödinger equation for the case \( E < V_0 \), one easily finds that the symmetric (even) wavefunctions can be written as
\begin{equation}
    \psi(x) = 
    \begin{cases}
    A \sin(k(x + b/2)), & -b/2 \leq x \leq -d/2, \\
    C \cosh(\kappa x), & -d/2 < x < d/2, \\
    -A \sin(k(x - b/2)), & d/2 \leq x \leq b/2,
    \end{cases}
    \label{eq:psi0}
\end{equation}
where \(k = \sqrt{2mE}/\hbar\), \(\kappa = \sqrt{2m(V_0 - E)}/\hbar\), or alternatively \(\kappa^2 = k_0^2 - k^2\) with \(k_0 = \sqrt{2mV_0}/\hbar\). Continuity of the wavefunction and its derivative  at \(x=\pm d/2\) leads to  the following relations:
\begin{equation}
    \frac{C}{A} = \frac{\sin(kL)}{\cosh(\kappa d/2)},
\end{equation}
and
\begin{equation}
    k \cot(kL) + \kappa \tanh(\kappa d/2) = 0 ,
\label{eq:E1}
\end{equation}
where \(L = (b-d)/2\). 
The transcendental equation (\ref{eq:E1}) thus determines the allowed energies (i.e., the wavevectors $k$) for the symmetric states. We note that for fixed $L$, the minimum barrier height for which a bound state still exists with $E<V_0$ is
\begin{equation}
V_{0,\min} = \frac{\hbar^2 \pi^2}{8 m L^2} \, ,
\end{equation}
corresponding to the situation  when the bound state energy $E$ just touches the top of the barrier. Similarly, for fixed $V_0$, the maximum barrier width that allows for a bound state with $E<V_0$ is
\begin{equation}
d_{\max}= b-\frac{\pi\hbar}{\sqrt{2mV_0}} \, .
\end{equation}

The solutions for the antisymmetric (odd) wavefunctions are of the form
\begin{equation}
    \psi(x) = 
\begin{cases}
A' \sin(k(x + b/2)), & -b/2 \leq x \leq -d/2, \\
C' \sinh(\kappa x), & -d/2 < x < d/2, \\
A' \sin(k(x - b/2)), & d/2 \leq x \leq b/2,
\end{cases}
\end{equation}
where
\begin{equation}
    \frac{C'}{A'}  =  \frac{\sin(kL)}{\sinh(\kappa d/2)}
\label{eq:A2}
\end{equation}
and 
\begin{equation}
     \kappa  \tan(kL) +  k \tanh(\kappa d/2) = 0 ,
\label{eq:E2}
\end{equation}
with the last equation determining the possible energies for odd states.

Note that for the case where \(E > V_0\) we can easily extend our previous solutions just making the change \(\kappa \to iq\), where \(q = \sqrt{2m(E-V_0)}/\hbar\). In this case, the exponential behavior of the solution inside the barrier (the region \(-d/2<x<d/2\)) gives place to an oscillatory solution, but we shall not pursue this analysis any further, for   here  we are mainly interested in the case \(E < V_0\), when the particle  can be said `to tunnel' through the barrier and move between the two wells, as discussed next.

\subsection{Tunneling Time for the Ground State}

The  mean tunneling time, $\bar{\tau}$,  for a particle in the ground state to traverse the barrier  
can now be obtained from  Eq.~(\ref{eq:tau_ab}), after setting $a=-b/2$, $x=-d/2$,  $b=d/2$, and using  the  symmetric wavefunction 
given in (\ref{eq:psi0}).
Performing the integrals in Eq.~(\ref{eq:tau_ab}), we  obtain after some simplification
\begin{equation}
\begin{split}
    \bar \tau =  \frac{2m}{\hbar}\frac{\tanh\left(\frac{\kappa d}{2}\right)}{\kappa} &\left[\frac{d}{2}+ L \left(1 - \frac{k_0^2}{2k^2}+ \frac{k_0^2 \cosh(\kappa d) }{2k^2} \right)
    \right.\\[4pt] &\left.
    + \frac{k_0^2 \sinh(\kappa d)}{2\kappa k^2} \right].
\end{split}
    \label{eq: tau2}
\end{equation}
This equation thus gives  the mean tunneling time  in terms of the ground-state energy $E$, determined by the lowest $k$ obtained from \eqref{eq:E1}, and the potential parameters $d$,  $L$, and $V_0$ (determined by $\kappa$).

In the limit of a high barrier, i.e.~\(V_0\gg E\),  so that \(\kappa d\gg 1\) and \(\kappa L\gg 1\),  one gets
\begin{equation}
    \bar \tau \approx \frac{mLk_0^2 }{2\hbar \kappa k^2 } e^{\kappa d} \approx \frac{m Lk_0}{2\hbar k^2} e^{k_0 d},
    \label{eq: tau-lim}
\end{equation}
where in the last passage we have set $\kappa\approx k_0$. The exponential factor in Eq.~\eqref{eq: tau-lim} shows the characteristic suppression of tunneling probability, and hence the increase in  tunneling time,  with the barrier width $d$.

\begin{figure}
    \centering
    \includegraphics[width=0.9\linewidth]{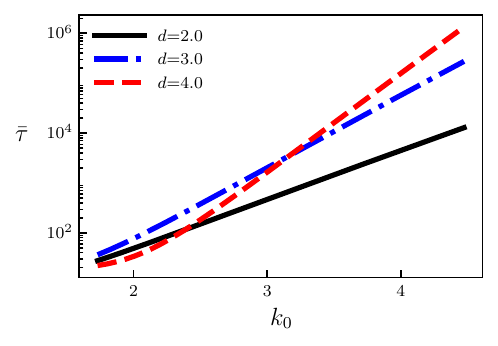}
    \caption{Mean tunneling time as a function of $k_0=\sqrt{2V_0}$  for three values of the barrier thickness $d=\{2.0,\, 3.0,\, 4.0\}$.
    Note the exponential growth with increasing $k_0$. Here quantities are in dimensionless units; see text.}
    \label{fig:tau_V0}
\end{figure}

\begin{figure}
    \centering
    \includegraphics[width=0.9\linewidth]{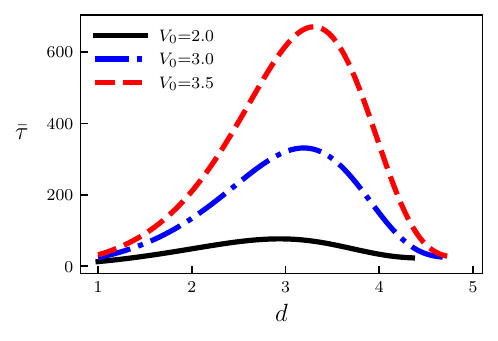}
    \caption{Mean tunneling time as a function of $d$ for three values of the barrier height $V_0=\{2.0,\, 3.0,\, 3.5\}$.
    Note that the mean tunneling time has a maximum. Here quantities are in the same dimensionless units as in Fig.~\ref{fig:tau_V0}.}
    \label{fig:tau_d}
\end{figure}

In Fig.~\ref{fig:tau_V0}, we present, in a semi-logarithmic plot, the behavior of $\bar{\tau}$ given by Eq.~\eqref{eq: tau2} as a function of $k_0$ (recall that $k_0 = \sqrt{2mV_0}/\hbar$) for three values of $d$.  
Here we plot $\bar{\tau}$ in dimensionless units, with $\hbar=m=1$ and $b=6$, for  $d = 2.0$, $3.0$, and $4.0$. For each set of parameters,  we solve the transcendental equation \eqref{eq:E1} to determine the lowest $k$, so we can compute $\bar{\tau}$ from Eq.~\eqref{eq: tau2}.  We see from this figure that $\bar{\tau}$ increases monotonically with $k_0$, as expected, and exhibits an exponential growth at large $k_0$, in agreement with Eq.~\eqref{eq: tau-lim}---a behavior clearly reflected by the linear asymptotic trend in the semi-logarithmic plot.
Figure \ref{fig:tau_d} shows $\bar\tau$ as a function of $d$, for three different values of $V_0$. One sees from this figure that, starting from a small value of $d$,  the mean tunneling time first increases with the barrier width, as expected, then reaches a maximum value,  and  decreases  after that. 
This behavior reflects a competition between two effects. On the one hand, increasing $d$ widens the barrier, which tends to suppress tunneling. On the other hand, reducing the classically allowed region (i.e., decreasing $L$) increases the kinetic energy inside the wells, causing the particle to traverse the allowed region more rapidly and encounter the barrier more frequently. In other words, the particle  `ricochets' ever so more rapidly between these two vertical  walls. Beyond a certain value of $d$ (at which $\bar\tau$ reaches a maximum), this confinement effect dominates, leading to a decrease in the tunneling time.  This interplay explains the overall shape of the curves shown in Fig.~3.
The fact that the function $\bar\tau(d)$, for fixed $V_0$, exhibits a maximum explains the curve crossings observed in Fig.~\ref{fig:tau_V0}. For example, in the case $V_0=2$, the maximum occurs near $d\simeq 3$ (see the black curve in Fig.~\ref{fig:tau_d}), which accounts for the observation that around $k_0=2$ in Fig.~\ref{fig:tau_V0} the curve corresponding to $d=3$ lies above those for $d=2$ and $d=4$. For sufficiently large $k_0$, however, the maxima of $\bar\tau(d)$ shift to values of $d$ beyond the three considered in Fig.~\ref{fig:tau_V0}, and the curves become monotonically ordered from bottom to top as $d$ increases.

\begin{figure}
    \centering
    \includegraphics[width=0.9\linewidth]{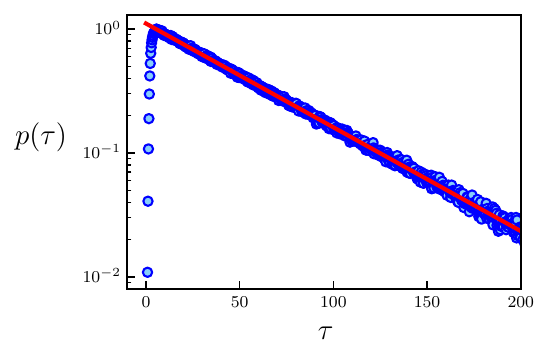}
    \caption{Tunneling time distribution (blue circles) for the double square well potential with dimensionless parameters $d=2.0$
    and $V_0=2.0$, computed from  $10^6$ trajectories. An exponential fit (red line),  $p(\tau) = C\exp(-\tau / \tau_l)$, where $\tau_l=51.80$, is also shown.}
        \label{fig:exp_decay_DSWP}
\end{figure}

In Fig.~\ref{fig:exp_decay_DSWP}, we present, on a semi-logarithmic scale, the empirical distribution of tunneling times, $p(\tau)$ (blue circles), obtained from $10^6$ trajectories generated by numerically integrating the stochastic differential equation~(\ref{eq:dx12}) using the Euler--Maruyama method with a time step  $\Delta t=10^{-4}$ and setting $d=2$ and $V_0=2$ (all quantities in dimensionless units). For each trajectory, we start the simulation with the particle   initially at $x = -(d+b)/4$ and  stop it when the particle reaches the point  $x = (d+b)/4$, where the starting and ending points correspond to the midpoints of the left and right wells, respectively. The traversal time  between these two points is then recorded for each   simulated trajectory  for subsequent statistical analysis. The exponential tail predicted by the theory discussed in Sec.~\ref{sec:dist} [see Eq.~(\ref{eq:exp})] is clearly recognized in Fig.~\ref{fig:exp_decay_DSWP} as the straight-line behavior (red line) in the semi-logarithmic plot. The mean tunneling time extracted from the simulations, $\bar{\tau}_{\mathrm{num}} = 54.45$, is in good agreement with the theoretical value $\bar{\tau}_{\mathrm{th}} = 55.90$ obtained from Eq.~\eqref{eq:tau_ab} for this case. (The small discrepancy stems from finite statistics.)

We briefly comment on the tunneling process from an energetic viewpoint. 
We recall that in SM the instantaneous energy of the particle is given by \cite{Nelson1979}
\begin{equation}
    {\cal E}(x,t)
    = \frac{m}{2}\!\left[v^2(x,t) + u^2(x,t)\right] + V(x),
\end{equation}
which for stationary  states (i.e., $v=0$) reduces to
\begin{equation}
    {\cal E}(x)
    = \frac{m}{2}\,u^2(x) + V(x),
    \label{eq:Ex}
\end{equation}
where $u(x)=(\hbar/2m)\psi'(x)/\psi(x)$.
For the present case of tunneling in the ground-state, the mean energy coincides with the ground-state eigenvalue:
\begin{equation}
   \ev{{\cal E}(x)}
   = \int {\cal E}(x)\,\rho(x)\,dx
   = E_0,
\end{equation}
where $\rho(x) = |\psi_0(x)|^2$ and $\psi_0(x)$ is the ground-state wavefunction.
%
Equation~\eqref{eq:Ex} shows that, within the barrier region, the instantaneous stochastic energy always satisfies ${\cal E}(x) \ge V_0$. 
Thus, strictly speaking, the standard textbook picture of quantum tunneling---where the particle traverses a classically forbidden region with insufficient energy---does not apply in the SM framework. 
Instead, the stochastic dynamics continually exchanges energy with the background noise, enabling the particle to transiently acquire enough energy to cross the barrier, even though the mean value $E_0$ remains below the barrier height.

\begin{figure}
    \centering
    \includegraphics[width=0.9\linewidth]{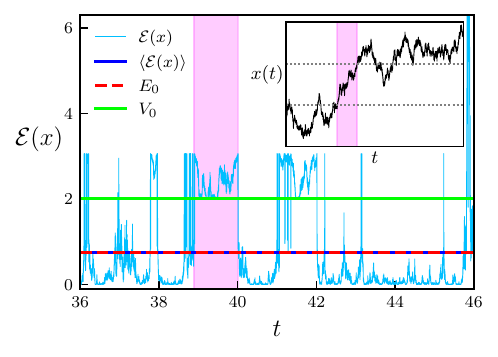}
    \caption{Sample time series of the particle’s instantaneous energy ${\cal E}(x)$ in the double square-well potential obtained via stochastic mechanics. The mean energy $\ev{{\cal E}(x)}$ coincides with the ground-state eigenenergy $E_0$ of the system. The inset displays the corresponding stochastic trajectory, with the barrier edges indicated by the dotted lines. The instantaneous energy exceeds the barrier height $V_0$ whenever the particle is inside the barrier region, see, e.g., the highlighted region. The instantaneous energy also exhibits a sharp rise whenever the particle tries to approach the infinite walls of the potential.
}
    \label{fig:Et}
\end{figure}

This mechanism is illustrated in Fig.~\ref{fig:Et}, which shows  the instantaneous total energy  for a segment of a representative trajectory exhibiting a tunneling event (highlighted region). Note that the time-averaged energy (blue solid line) closely matches the ground-state value $E_0$ (red dashed line), confirming the consistency between the stochastic simulations and the stationary quantum solution. The inset shows the corresponding stochastic path, with the barrier edges indicated by dotted horizontal lines. Comparing the inset with the main plot, one sees that whenever the particle `penetrates' the barrier from either side, its instantaneous energy rises above $V_0$, in agreement with the preceding discussion.
The instantaneous energy also exhibits sharp spikes whenever the particle approaches the infinite walls, thereby dynamically repelling the particle from  these regions.

\subsection{Quantum Mechanics Prediction: Probability Oscillation Period}

In quantum mechanics, tunneling through a barrier in a double-well potential is commonly analyzed within the two-state approximation. In this framework, one defines the localized states  
\[
\psi_{L,R}(x) = \frac{1}{\sqrt{2}}\,[\psi_0(x) \pm \psi_1(x)],
\]
where $\psi_0(x)$ and $\psi_1(x)$ denote the ground and first excited eigenstates, respectively. The wave functions $\psi_L(x)$ and $\psi_R(x)$ are then predominantly localized in the left and right wells. If the system is initially prepared, say, in the left state, $\Psi(x,0) = \psi_L(x)$, the probability density $p(x,t) = |\Psi(x,t)|^2$ oscillates periodically in time with period 
\begin{equation}
    T = \frac{2\pi\hbar}{\Delta E},
    \label{eq:T}
\end{equation}
where $\Delta E = E_1 - E_0$ is the energy difference between the first excited and ground states.
The quantity $\Delta E$ is usually interpreted as the \textit{energy splitting} of the single-well ground-state level due to the coupling between the two wells. Within this picture, one can regard the tunneling time predicted by quantum mechanics as  
\[
\tau_{\rm QM} = \frac{T}{2}.
\]
It is important to note, however, that within the standard interpretation of quantum mechanics, it is not meaningful to speak of ``tunneling'' when the particle is in an energy eigenstate---such as the ground state---since the corresponding probability density is stationary. In that case, one can only say that the particle is equally likely to be found in either well.

The situation is fundamentally different in Nelson's stochastic mechanics, where the probability density arises from an underlying stochastic process. In this framework, a \textit{mean tunneling time}, $\bar{\tau}$, can be computed even for stationary states. Despite this conceptual difference, we shall show below that there exists a direct quantitative relation between the mean tunneling time predicted by the stochastic mechanics formalism and that inferred from the two-state quantum-mechanical description. 

To proceed, let us analyze the oscillation period $T$ in the limit $V_0 \gg E_1 > E_0$. For this purpose, we first evaluate the energy difference $\Delta E = E_1 - E_0$ in this limit. Introducing the dimensionless variables $z_1 = k_1L$ and $z_2 = k_2L$ in Eqs.~\eqref{eq:E1} and \eqref{eq:E2}, where $k_1$ and $k_2$ denote the wavenumbers associated with the ground and first excited states, respectively, we obtain  
\begin{equation}
     \tan(z_1) + \frac{k_1}{\kappa_1} \coth\!\left(\frac{\kappa_1 d}{2}\right) = 0,
     \label{eq:E1_z}
\end{equation}
and  
\begin{equation}
      \tan(z_2) + \frac{k_2}{\kappa_2} \tanh\!\left(\frac{\kappa_2 d}{2}\right) = 0.
      \label{eq:E2_z}
\end{equation}
For \(V_0 \to \infty\) the energies \(E_1\) and \(E_0\) coalesce to the ground-state energy, $E_{\infty} = \hbar^2\pi^2/2mL^2$ (or \(k = \pi/L\)), of a single infinite well of width $L$, so that \(z \to \pi\) in this limit. Making appropriate   approximations in Eqs.~\eqref{eq:E1} and \eqref{eq:E2}, one obtains
\begin{align}
    z_1 - \pi &= -\frac{k}{\kappa}(1 + 2e^{-\kappa d})  
    \label{eq:z1} , \\
    z_2 - \pi &= -\frac{k}{\kappa}(1 - 2e^{-\kappa d}),
    \label{eq:z2}
\end{align}
where we have set \(k_2 \approx k_1 = k\) and \(\kappa_2 \approx \kappa_1 = \kappa\). Subtracting (\ref{eq:z1}) from (\ref{eq:z2}), we find
\begin{equation}
    \Delta k = k_2 - k_1 = \frac{z_2 - z_1}{L} = \frac{4k}{L\kappa}e^{-\kappa d},
\end{equation}
from which follows that 
\begin{equation}
\Delta E \approx \frac{4\hbar^2k^2}{L\kappa m}e^{-\kappa d}.
\label{eq: Delta E}
\end{equation}
%
In the limit of a very high barrier the oscillation period \(T\), see Eq.~\eqref{eq:T}, then becomes
\begin{equation}
    T \approx \pi \frac{L\kappa m}{2\hbar k^2}e^{\kappa d}
    \label{eq:TDSW}
\end{equation}
Comparing Eqs.~\eqref{eq: tau-lim} and  \eqref{eq:TDSW} then yields the relation
\begin{equation}
    \frac{\tau_{\rm QM}}{\bar\tau} 
    = \frac{\pi}{2},
    \label{eq:T/2tau}
\end{equation}
as already anticipated in the introduction; see Eq.~\eqref{eq:1}. As we will see next, relation \eqref{eq:T/2tau}, derived above for the double square well, is valid in general for any double-well potential in the high-barrier limit.


\section{Tunneling Time for Generic Double-Well Potentials: WKB Analysis}

\label{sec:wkb}



\begin{figure}
    \centering
    \includegraphics[width=0.9\linewidth]{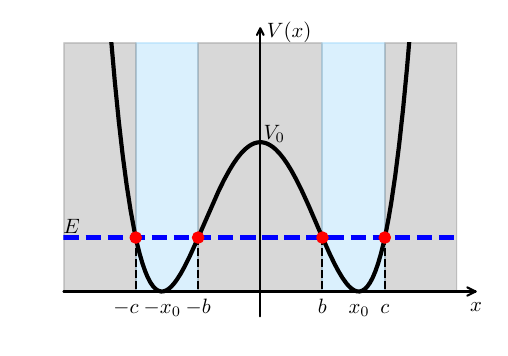}
    \caption{Symmetric double-well  potential with barrier height $V_0$ and minimum point at $\pm x_0$. The dashed blue line represents the particle energy $E$. The red points located at $\pm c$ and $\pm b$ are the classical turning points. The blue areas correspond to the classically allowed regions and the gray areas correspond to the classically forbidden regions. Classically, a particle with energy $E$ will oscillate between $b$ and $c$ (or equivalently, between $-c$ and $-b$) and will never be found in the forbidden regions. In quantum mechanics, the particle can be in the interval $[-c, -b]$ as well as in the interval $[b, c]$ with equal probability and can travel from one interval to another through the forbidden region $[-b, b]$, this is the tunnel effect.}
    \label{fig:DWP}
\end{figure}

For simplicity (but without loss of generality),  we consider a symmetric double-well potential, $V(x) = V(-x)$, with two minima at $x=\pm x_0$, i.e., $V'(\pm x_0)=V(\pm x_0)=0$, and a local maximum (barrier) at the origin, i.e., $V'(0)=0$ and $V(0)=V_0>0$. For a particle in a state of energy $E$, the classical turning points are denoted by $\pm c$ and $\pm b$, where $V(\pm c) = V(\pm b) = E$.  The classically allowed regions are therefore the intervals $[-c,-b]$  and  $[b,c]$ in the left and right  wells of the potential, respectively, whereas the interval $[-b,b]$ corresponds to the classically forbidden region (inside the barrier); see Fig.~\ref{fig:DWP}.
We assume furthermore that $V_0$ is sufficiently high, so that  $E_0, E_1<V_0$, where, as before, $E_0$ and $E_1$ denote the energies of ground state and first excited state, respectively. 

For a continuous potential, it is not straightforward to identify where the particle begins and ends the tunneling process, since the barrier does not necessarily have sharp boundaries. Alternative approaches based on the construction of time operators, such as that of Ref.~\cite{DiasParisioPRA2017}, address this issue in a different manner by defining a traveling time between prescribed spatial regions---whether or not tunneling occurs---through the expectation value of an appropriate self-adjoint operator \cite{deLaraBeims2024_TravelingTime}.
Within the SM approach, Köppe \textit{et al}.~\cite{KoeppeJMP2018_QHE_1D} suggested instead considering the traversal time for a particle initially released at a position $x$ in the left well (i.e., $-\infty < x < -b$) to reach the right classical turning point at $x=b$, and then averaging over the initial positions. 
However, for a high barrier, i.e., $V_0\gg E_0$, the barrier walls become sufficiently sharp that one can assume, as was the case of the double square well potential, that tunneling effectively starts when the particle reaches the right classical turning point, $x=-b$, of the left well and ends when the particle `emerges' at the symmetric point, $x=b$, in the right well; see Fig.~\ref{fig:DWP}.  Also, because the potential is assumed to have unbounded support, we take $a=-\infty$, corresponding to a reflecting boundary at minus infinity; see Sec.~\ref{sec:dist}. With these choices, then according to Eq.~\eqref{eq:tau_ab} the mean tunneling time is given  by

\begin{align}
   \bar\tau &=\frac{2m}{\hbar}\int_{-b}^b \frac{1}{p(x)}\left(\int_{-\infty}^x p(y) dy\right) dx  ,
   \label{eq:tau_infty}
\end{align}
where $p(x)=|\psi(x)|^{2}$ is the probability density of the ground (symmetric) state.
Now, noting that  the probabilities of finding the particle in the classically forbidden regions to the left of the potential well, i.e., for $x<-c$, and inside the barrier, i.e., for $|x|<b$,  are negligible  in the high-barrier limit, we can  simplify Eq.~\eqref{eq:tau_infty} to 
\begin{equation}
\bar{\tau}=\frac{2m}{\hbar}\left[\int_{-b}^{b}\frac{dx}{p(x)}\right]\left[\int_{-c}^{-b}p(y)dy\right].
\label{Tau_z2}
\end{equation}
We thus see that, to estimate $\bar{\tau}$, it suffices to know the wavefunction in the classically allowed region, $-c < x < -b$, within the left well, and in the classically forbidden region, $-b < x < b$, under the barrier. This can readily be achieved within the Wentzel--Kramers--Brillouin (WKB) approximation. 

We recall that the  WKB approximation \cite{LandauLifshitzQM,MerzbacherQM} is valid when the potential changes very slowly compared to the particle wavelength, which implies  the condition:
\begin{equation}
    \frac{\hbar^{2}}{2m}\,\frac{|V'(x)|}{|E-V(x)|^{3/2}} \ll 1.
    \label{WKB_cond}
\end{equation}
We shall suppose throughout this section that this condition is satisfied by our generic potential $V(x)$ in the regions of interest.  Let us now introduce a few standard quantities used in the WKB formalism, namely the local wave number,
\begin{align}
k(x) &= \frac{\sqrt{2m[E - V(x)]}}{\hbar}, \qquad E > V(x),
\end{align}
the attenuation coefficient (or decay constant),
\begin{align}
\kappa(x) &= \frac{\sqrt{2m[V(x) - E]}}{\hbar}, \qquad V(x) > E,
\end{align}
and the barrier integral,
\begin{equation}
\Phi = \int_{-b}^{b} \kappa(x)\,dx.
\label{eq:Phi}
\end{equation}

Using the WKB connection formulas for matching wave functions across turning points~\cite{LandauLifshitzQM,MerzbacherQM}, the symmetric wave function can be written in the relevant regions as  
\begin{align}
\psi(x) \;\approx
\begin{cases}
\dfrac{A}{\sqrt{k(x)}} \cos\!\left(\displaystyle\int_{-b}^{x} k(x^{\prime})\,dx^{\prime} + \dfrac{\pi}{4}\right), & -c < x < -b, \\[1.2em]
\dfrac{B}{\sqrt{\kappa(x)}} \cosh\!\left(\displaystyle\int_{0}^{x} \kappa(x^{\prime})\,dx^{\prime}\right), & -b < x < b, \label{eq:psi_wkb}\\[1.2em]
\dfrac{A}{\sqrt{k(x)}} \cos\!\left(\displaystyle\int_{b}^{x} k(x^{\prime})\,dx^{\prime} - \dfrac{\pi}{4}\right), & b < x < c,
\end{cases}
\end{align}
where $A = B\,\exp\!\left(\tfrac{\Phi}{2}\right)$, with $\Phi$ defined in Eq.~\eqref{eq:Phi}.  
The constant $A$ is determined by the normalization condition $\int_{-c}^{c} |\psi(x)|^2\,dx \approx 1$.  
The antisymmetric (odd) wave function has the same form, except that $\cosh$ is replaced by $\sinh$ in the barrier region.

Neglecting the probability of finding the particle inside the barrier and noting that the probability distribution is equal in both wells, we can write 
\begin{align}
  \int_{-c}^{-b}p(x)dx=  \int_{b}^{c}p(x)dx\approx 1/2.
  \label{eq:int1/2}
\end{align}
We also have  
\begin{align}
   \int_{b}^{c} p(x)\,dx &= \int_{b}^{c} \frac{A^2}{k(x)} 
   \cos^2\!\left(\int_{b}^{x} k(x')\,dx' - \frac{\pi}{4}\right) dx \cr
   &\approx \frac{A^2}{2} \int_{b}^{c} \frac{dx}{k(x)},
\end{align}
where we have replaced the rapidly oscillating term $\cos^2(\cdots)$ by its average value of $1/2$, consistent with the WKB condition given in Eq.~\eqref{WKB_cond}.
Let us now define the classical oscillation period in the classically allowed regions as  twice the time to go from one classical turning point to the other within a potential well:
\begin{equation}
T_{\text{cl}} = 2 \int_b^c \frac{dx}{v_{\rm cl}(x)} = \frac{2m}{\hbar} \int_b^c \frac{dx}{k(x)},
\label{eq:T_classical}
\end{equation}
where $v_{\rm cl}(x) = \hbar k(x)/m$ denotes the classical velocity. Using this definition, we can write  
\begin{align}
   \int_{b}^{c} p(x)\,dx \approx \frac{A^2}{4m}\, \hbar T_{\text{cl}},
\end{align}
which, in view of Eq.~\eqref{eq:int1/2}, yields  
\begin{equation}
A^2 \approx \frac{2m}{\hbar T_{\text{cl}}},
\end{equation}
and consequently,  
\begin{equation}
B^{2} \approx \frac{2m}{\hbar T_{\text{cl}}} e^{-\Phi}.
\label{eq:B2}
\end{equation}

Having obtained the WKB wavefunctions, we can now proceed to compute the mean tunneling time.  Using Eqs.~\eqref{eq:psi_wkb} and \eqref{eq:int1/2}  in  Eq.~\eqref{Tau_z2}, one gets
\begin{equation}
\bar \tau \approx\frac{m}{\hbar B^{2}}\int_{-b}^{b}\frac{\kappa(x)}{\cosh^{2}(\int_{0}^{x}\kappa(x^{\prime})dx^{\prime})}dx.
\end{equation}
Setting $\kappa(x)\approx\kappa_{0}$, where $\kappa_{0}$ denotes the average value,  so that $\int_{0}^{x}\kappa(x^{\prime})dx^{\prime}\approx\kappa_{0}x$, and performing the integral above, we find in the high-barrier limit ($\kappa_{0}b\gg 1$):
\begin{equation}
\bar \tau \approx \frac{2m \tanh(\kappa_0b)}{\hbar B^2}\approx \frac{2m }{\hbar B^2} .
\end{equation}
Using Eq.~\eqref{eq:B2}, we then obtain the WKB estimate for $\bar \tau$:
\begin{equation}
\bar \tau\approx e^{\Phi} T_{\text{cl}} .
\label{eq:tau_wkb}
\end{equation}

Let us now turn to the probability oscillation period in the framework of quantum mechanics.
It is well known \cite{LandauLifshitzQM,David-Park} that in the WKB approximation the energy splitting for a symmetric double-well in the high barrier limit ($\Phi \gg 1$) is given by 
\begin{equation}
\Delta E \approx \frac{2\hbar}{T_{\text{cl}}}  e^{-\Phi}.
\end{equation}
Inserting this into Eq.~\eqref{eq:T}, we get
\begin{equation}
T  \approx  \pi e^{\Phi} T_{\text{cl}} .
\label{T_wkb}
\end{equation}

It is thus noteworthy that both the mean tunneling time  and the quantum-probability oscillation period  have the same dependence on the barrier integral $\Phi$ and on the classical time $T_{\text{cl}}$, differing only by a numerical factor; see Eqs.~\eqref{eq:tau_wkb} and \eqref{T_wkb}. From these relations we then immediately recover relation \eqref{eq:T/2tau} for the  ratio (in the high-barrier limit) between the  tunneling time, $\tau_{\rm QM}=T/2$, estimated from quantum mechanics and the mean tunneling time, $\bar\tau$, predicted by stochastic mechanics. 
Note, however, that while Eq.~\eqref{eq:T/2tau} was derived for the double square well, the WKB calculation presented above is valid  regardless of the specific shape of the potential  or energy level in the WKB limit.
Both calculations yield the same numerical factor between $\tau_{\rm QM}$ and $\bar{\tau}$; yet it remains unclear whether the value $\pi/2$ admits a simple physical interpretation.

\section{Application to Ammonia}

\label{sec:RM}


A well-known example of quantum tunneling is the inversion of the ammonia molecule (NH$_3$). In this process, the nitrogen atom can penetrate the potential barrier formed by the plane of the three hydrogen atoms, effectively oscillating between two equivalent configurations on either side of the plane. Within quantum mechanics, this inversion is understood as the time evolution of a localized state that arises from the coherent superposition of the system’s symmetric (even) and antisymmetric (odd) energy eigenstates.
The transition frequency---here denoted by $\nu_{\mathrm{QM}}$---for such localized states is related to the energy splitting $\Delta E$ between the ground and first excited states of ammonia, or more exactly $\nu_{\rm QM} ={\Delta E}/{2\pi\hbar}$. The  experimental value for the ammonia inversion frequency is
approximately $24\,\text{GHz}$ \cite{TownesSchawlow1955}. 

In this section, we shall compute the mean tunneling time $\bar\tau$ predicted by Nelson's stochastic mechanics for the ground state of the ammonia molecule by using suitable double-well potentials. Using the correspondence given in Eq.~\eqref{eq:T/2tau} between the mean tunneling time, $\bar \tau$, obtained from stochastic mechanics and  the oscillation time, $T$, computed from quantum  mechanics, we can then write a stochastic-mechanics prediction for the ammonia inversion frequency as follows:
\begin{equation}
   \nu_{\rm SM} \equiv  \frac{1}{\pi \bar\tau}.
    \label{eq:nu_SQ}
\end{equation}
As argued above,  $\nu_{\rm SM}$ should in general be relatively close to the inversion frequency,  $\nu_{\rm QM}$, predicted by quantum mechanics.
Indeed, we will see below that the prediction for the inversion frequency defined in Eq.~\eqref{eq:nu_SQ} agrees remarkably well with the experimental value for the ammonia inversion frequency, thus validating the calculation of the mean tunneling time within the stochastic mechanics formalism.

To investigate the tunneling dynamics of the ammonia molecule within the framework of stochastic mechanics, we model the system as a particle of mass $m$---the reduced mass of the molecule---moving in a suitable symmetric double-well potential. Several potential models have been proposed to describe the inversion motion in ammonia, including the Dennison--Uhlenbeck potential~\cite{DennisonUhlenbeck1932}, the Rosen--Morse potential~\cite{RosenMorse}, and the Manning potential~\cite{Manning},  among others~\cite{Peacock-Lopez,Sitnitsky2017}. However, as discussed above, the near identity between the tunneling frequency predicted by stochastic mechanics, $\nu_{\rm SM}$, and its quantum-mechanical counterpart, $\nu_{\rm QM}$, is expected to hold irrespective of the detailed form of the potential (in the high-barrier limit). Hence, for our purposes it suffices to consider a representative potential model for the ammonia molecule. In what follows, we therefore adopt the double Rosen--Morse potential (RMP), which has been widely used to model the ammonia inversion dynamics~\cite{RosenMorse,YangHan_IJMS_2021_Ammonia}.

\begin{figure}
    \centering
    \includegraphics[width=0.9\linewidth]{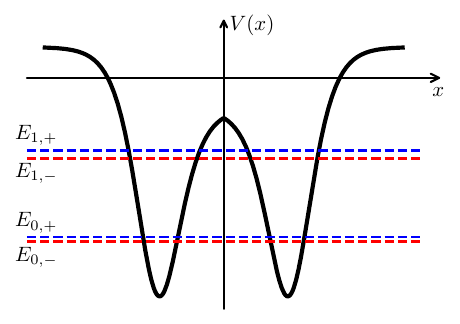}
    \caption{Double Rosen-Morse potential defined by expression \eqref{eq: DRMP}. The first two pairs of energy levels are shown. The red lines are the energies of even states and the blue lines are the energies of odd states.}
    \label{fig: DRMP}
\end{figure}

The double Rosen-Morse potential was introduced by Rosen and Morse \cite{RosenMorse} to model the vibration of polyatomic molecules, with special attention to the ammonia inversion problem, which is the prototypical example of this kind of problem. 
It has the following form (see Fig.~\ref{fig: DRMP}):
\begin{equation}
    V(x) = 
    \begin{cases}
        A \tanh\left(\dfrac{x}{d} - k\right) - B\sech^2 \left(\dfrac{x}{d} - k\right), & x\geq 0,\\[4pt]
        -A \tanh\left(\dfrac{x}{d} + k\right) - B\sech^2 \left(\dfrac{x}{d} + k\right), & x\leq 0,
    \end{cases}
    \label{eq: DRMP}
\end{equation}
where $A$, $B$, $d$ and $k$ are adjustable parameters. 
The reduced mass of the nitrogen atom relative to the hydrogen atoms is
\begin{equation}
    m = \frac{3m_Hm_N}{3m_H + m_N}
    \label{eq: mass}
\end{equation}
where $m_H$ and $m_N$ are the masses of the hydrogen and nitrogen atoms, respectively. 

To choose the potential parameters one typically uses three experimental data \cite{DennisonUhlenbeck1932}, namely: the energy separation, $E_{1,+}-E_{0,-}= 950\text{ cm}^{-1}$, between the two lowest pair of levels and the 
separations within each pair of levels, $\Delta E_0 = E_{0,+}-E_{0,-}=0.8\text{ cm}^{-1}$ and $\Delta E_1=E_{1,+}-E_{1,-}= 33\text{ cm}^{-1}$ (where $1\text{ cm}^{-1} = 1\,hc \text{ J} \approx 1.988\cdot 10^{-23}\text{ J}$), see Fig.~\ref{fig: DRMP}. Since there are four parameters and only three experimental data to fit, the best that can be done in this case is to determine a range of values for each parameter based on the reasonable shape of the potential, as described in \cite{RosenMorse}.
These restrictions provide the following ranges of possible values for the RMP parameters: $
0< A<1000 \text{ cm}^{-1}, \ 2200 \text{ cm}^{-1} < B<  3000 \text{ cm}^{-1}, \
0.16 \text{ \AA}< d<  0.185 \text{ \AA}$, and $2.20 < k<  2.24$.

From Eq.~\eqref{eq: DRMP} one finds that the two potential minima are located at $x=\pm x_0$, where
\begin{equation}
    x_0 = kd - \tanh^{-1}\left(\frac{A}{2B}\right)d.
\end{equation}
The barrier  height, $V_0 = V(0) - V(x_0)$, is given by
\begin{equation}
    V_0 = \frac{A^2}{2B} - A\tanh k + B\tanh^2 k,
\end{equation}
whilst the dissociation energy (or well depth), $V_D$, corresponding to the difference between the potential asymptotic value, $V(x\to\infty)=A$, and its minimum, $V(x_0)$,   reads
\begin{equation}
V_D =   A + B + \frac{A^{2}}{4B}.
\label{eq:VD}
\end{equation}

Within the ranges of allowed parameters mentioned above, one can verify that the values of $x_0$ and $V_0$  differ at most by eight percent. So, any form of the potential \eqref{eq: DRMP} within the acceptable range of parameters will have $x_0$ near 0.38 {\AA} and $V_0$ near 2050 cm$^{-1}$.
%
With these considerations in mind, we choose the values of the parameters in order to get a value of $\Delta E_0$ very close to the experimental value of $0.8 \text{ cm}^{-1}$, since this is the energy splitting that determines the quantum-mechanical tunneling frequency.  To do so, we compute numerically the energy eigenvalues of the RMP with the Numerov method \cite{Numerov1924} and apply a fit procedure  to determine the optimal RMP parameters, where we fix $d = 0.17 \text{ \AA}$ and $k=2.22$, and vary $A$ and $B$. The values obtained are as follows: $A = 398\text{ cm}^{-1}$ and $B = 2810 \text{ cm}^{-1}$,  which of course are  within the allowed ranges, as expected.

Having determined the relevant parameter values for the double Rosen-Morse potential, we can proceed to compute the mean tunneling time predicted by stochastic mechanics for the ground state of the ammonia molecule.
The classical turning points are $x=\pm b$ and $x=\pm c$, where $V(\pm c) = V(\pm b) = E_{0,-}$.
The ground state wave function, $\psi_0(x)$, and its energy, $E_{0,-}$, were calculated numerically using the Numerov method. 
We then numerically performed the integrations in Eq.~\eqref{eq:tau_infty} to obtain the mean tunneling time, $\bar \tau$, for a particle with  reduced mass $m$ given by Eq.~\eqref{eq: mass} in the ground state of the RMP.
We find  $\bar \tau = 13.4578$ ps, which in view of Eq.~(\ref{eq:nu_SQ}) corresponds to a tunneling frequency 
$\nu_{\rm SM}=23.65$  GHz,  which is indeed very close to the experimental value of 23.79 GHz \cite{TownesSchawlow1955}. This agreement is quite remarkable, when considering that the `conversion formula' \eqref{eq:T/2tau} between tunneling time and quantum oscillation period,  which led to Eq.~\eqref{eq:nu_SQ},  was derived in the asymptotic limit $V_0 \gg E$. Our results thus suggest that this correspondence is expected to hold for any practical tunneling situation where the quantum mechanical two-state approximation is valid, as in the inversion dynamics of  the ammonia molecule.

\begin{figure}
    \centering
\includegraphics[width=0.9\linewidth]{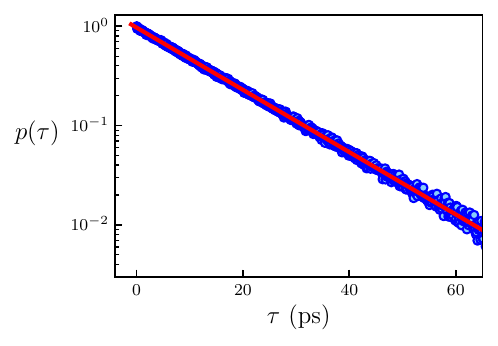}
\caption{Tunneling time distribution (blue circles) for the double Rosen-Morse potential with parameters $A=398 \text{~cm}^{-1}$, $B=2810 \text{~cm}^{-1}$, $k=2.22$ and $d=17\text{~pm}$, computed from $10^6$ trajectories.  An exponential fit (red line), $p(\tau) = C\exp[- (\tau / \tau_l)]$ with $\tau_l = 13.84 \text{~ps}$,  adjusts remarkably well the entire dataset.}1
    \label{fig:RMP_exp}
\end{figure}

In Fig.~\ref{fig:RMP_exp} we  show the distribution of tunneling times for the ground state of the RMP  with the parameters specified above, obtained from simulations of $10^6$ trajectories of the Brownian particle evolving under the corresponding stochastic mechanics diffusion process. The simulations were carried out using the Euler--Maruyama method with a time step of $10^{-3}$ in dimensionless units (or approximately $10^{-5}\,\text{ps}$). For each trajectory, the particle was initialized at $t=0$ at the right classical turning point of the left potential well, and the simulation was terminated when the particle reached the left classical turning point of the right well. The elapsed time between these two events defines the tunneling time for the trajectory. As seen in Fig.~\ref{fig:RMP_exp}, the resulting distribution again exhibits  a clear exponential decay, in agreement with the theoretical prediction discussed in Sec.~\ref{sec:dist}.
Note also that, in contrast to the corresponding distribution for the double square well (see Fig.~\ref{fig:exp_decay_DSWP}), the RMP case exhibits a much sharper rise from $p(0)=0$ to the exponential curve for $\tau>0$, with no events outside the exponential region detected within our set of $10^6$ simulations. Consequently, an exponential fit (red line) provides an excellent description of the entire dataset. The theoretical mean tunneling time, computed from Eq.~\eqref{eq:tau_infty}, is $\bar{\tau}_{\rm th}=13.46~\text{ps}$, in very good agreement with the corresponding numerical value, $\bar{\tau}_{\rm num}=13.86~\text{ps}$.

\section{Discussion}

In this work, we have employed Nelson’s stochastic mechanics (SM) to investigate tunneling-time statistics for bound states in symmetric double-well potentials. By combining the stochastic-trajectory framework with the theory of first-passage times, we obtained closed expressions for both the mean tunneling time, $\bar{\tau}$ and the asymptotic (tail) behavior of  the probability distribution $p(\tau)$. For the double square-well potential, an analytic formula was derived for $\bar{\tau}$ and shown to be in excellent agreement with large-scale numerical simulations of the underlying diffusion process.

A central result of this study is the universal (i.e., shape-independent) relation given in Eq.~\eqref{eq:1}, which links the mean stochastic-mechanical tunneling time $\bar{\tau}$ to the quantum-mechanical tunneling time $\tau_{\mathrm{QM}}$, defined as half the oscillation period of the probability of finding the particle in either well. This proportionality, derived analytically for the square double-well and confirmed through a WKB analysis for arbitrary smooth double-well potentials, establishes a precise correspondence between the stochastic and conventional quantum descriptions in the high-barrier limit. It should be noted, however, that in general the proportionality between $\bar{\tau}$ and $\tau_{\mathrm{QM}}$ may depend somewhat on the potential shape and on the specific stopping rule adopted to define the tunneling time.

\begin{figure}[t]
    \centering
    \includegraphics[width=0.9\linewidth]{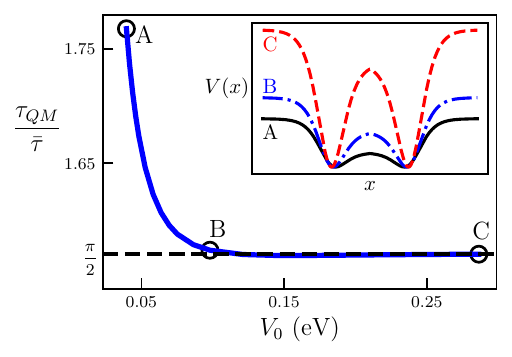}
    \caption{Ratio $\tau_{\mathrm{QM}}/\bar{\tau}$ between the quantum-mechanical tunneling time and the mean stochastic tunneling time as a function of the barrier height $V_0$ for the double Rosen–Morse potential. The parameters are fixed at $A = 398~\mathrm{cm^{-1}}$, $k = 2.22$, and $d = 17~\mathrm{pm}$, while $B$ is varied in the range $680~\mathrm{cm^{-1}} < B < 2810~\mathrm{cm^{-1}}$. The inset displays the  potential profiles corresponding to points A, B, and C indicated on the main curve, where the  barrier heights are 39.5 meV (A), 98.0 meV (B), and 286.5 meV (C), which correspond to $28.8\%$, $48.4\%$, and $71.6\%$ of the well depth $V_D$, respectively.}
    \label{fig:tau_V0_1}
\end{figure}

\begin{figure}
    \centering
    \includegraphics[width=0.9\linewidth]{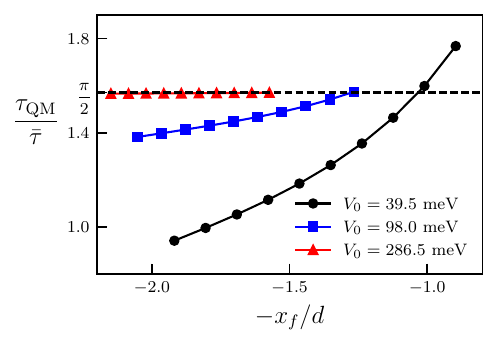}
    \caption{Ratio $\tau_{\rm QM}/\bar\tau$ as a function of $x_f/d$ for the Rosen-Morse potential, Eq.~\eqref{eq: DRMP}, with $V_0 = \{39.5\text{ meV},\, 98.0 \text{ meV},\, 286.5 \text{ meV}\}$ corresponding to points A, B and C in Fig.~\ref{fig:tau_V0_1}. The $x_f$ parameter, which delimits the tunneling region, varies from the classical turning point on the barrier ($b$) to the minimum potential point ($x_0$). If $V_0$ is high, then the ratio gets close to the value $\pi/2$ despite the choice of the parameter $x_f$.}
    \label{fig:T_2tau_DWP}
\end{figure}

The dependence of $\bar{\tau}$ on the potential shape is illustrated in Fig.~\ref{fig:tau_V0_1}, which shows the ratio $\tau_{\mathrm{QM}}/\bar{\tau}$ as a function of the barrier height $V_0$ for the Rosen–Morse potential. Here, the parameters are fixed at $A = 398~\mathrm{cm^{-1}}$, $k = 2.22$, and $d = 17~\mathrm{pm}$, while $B$ is varied in the range $680~\mathrm{cm^{-1}} < B < 2810~\mathrm{cm^{-1}}$. Remarkably, the prefactor approaches the universal value $\pi/2$ already at moderate values of the barrier height, as shown in Fig.~\ref{fig:tau_V0_1} (see, e.g., point B). We note that point C in this figure refers to the optimal RMP parameter values for the ammonia data  discussed above. For this point the barrier height, $V_0$, is approximately 72$\%$ of the well depth, $V_D$,  which indeed represents  a relatively high barrier, as seen in the  inset of Fig.~\ref{fig:tau_V0_1}.

Let us now briefly examine how the mean tunneling time, $\bar{\tau}$, depends on the stopping rule used to define the tunneling process. In general, the tunneling time may be defined as the duration required for the particle to move from some point $-x_f$ in the left well, say, in the classically allowed region  $-c < x_f < -b$, 
to the symmetric point $x_f$ in the right well. In this case, the mean tunneling time is given by a modified form of Eq.~\eqref{eq:tau_infty}, namely,  
\begin{align}
   \bar{\tau}(x_f) = \frac{2m}{\hbar} \int_{-x_f}^{x_f} \frac{1}{p(x)} \left( \int_{-\infty}^{x} p(y)\,dy \right) dx .
   \label{eq:tau_xf}
\end{align}
In Fig.~\ref{fig:T_2tau_DWP}  we plot the ratio $\tau_{\rm QM}/\bar \tau$ as a function of the initial tunneling point, $-x_f$, for the RMP potential, using parameters corresponding to points A, B, and C in Fig.~\ref{fig:tau_V0_1}. For the three curves shown, $-x_f$ is varied from the minimum point of the left well ($-x_0$) to the right classical turning point ($-b$). Since $\tau_{\mathrm{QM}}$ is fixed for each curve, one sees from Fig.~\ref{fig:T_2tau_DWP} that $\bar{\tau}$ decreases as $x_f$ decreases,  as expected, since the distance traveled within the tunneling region $(-x_f,x_f)$ becomes smaller. Also as expected, for a low barrier (black circles), the ratio $\tau_{\mathrm{QM}}/\bar{\tau}$ changes more significantly with $x_f$. As the barrier height increases—see, for instance, the upper curve in Fig.~\ref{fig:T_2tau_DWP}—this ratio becomes insensitive to the precise definition of the tunneling stopping rule, thereby confirming the universal character of the factor $\pi/2$ in this limit. This result provides further evidence that the classical turning points on both sides of the barrier can be safely used as reference positions in defining the quantum tunneling process, as argued above.

Physically, the proportionality between $\tau_{\mathrm{QM}}$ and $\bar{\tau}$  indicates that the diffusive motion intrinsic to stochastic mechanics reproduces the same characteristic timescale that, in quantum mechanics, governs coherent oscillations between the two localized states of the double-well. However, the mean tunneling time in SM is computed solely within the ground state and requires no prior knowledge of the energy splitting---a quantity that is essential for describing tunneling in standard quantum mechanics. The fact that $\tau_{\mathrm{QM}}$ exceeds $\bar{\tau}$ may be understood as reflecting the distinction between two complementary descriptions of the tunneling process. In the stochastic picture, $\bar{\tau}$ represents the mean first-passage time for an individual trajectory to cross the barrier from one well to the other, whereas in quantum mechanics $\tau_{\mathrm{QM}}$ relates to the collective oscillation period of the entire probability distribution. The latter entails a coherent redistribution of probability density between the wells---a process inherently slower than the stochastic transition of single trajectories. Within the high-barrier, two-level approximation, this interpretation provides a consistent physical link between single-trajectory diffusion dynamics and the collective evolution of quantum probability. (Whether a physical explanation exists for the emergence of $\pi/2$ as the universal ratio remains unclear at present.) 

As discussed above, the SM formulation offers several conceptual advantages for studying dynamical properties of quantum systems. Because SM is expressed directly in real space and time, it allows one to define operational quantities---such as traversal-time distributions---that are difficult or ambiguous within the standard Hilbert-space framework, where time appears only as an external parameter. The emergence of exponentially distributed tunneling times, derived analytically in general and confirmed numerically for both the square and Rosen–Morse double-wells, indicates that stochastic trajectories effectively sample barrier-crossing events as a Poisson-like process characterized by a single mean scale $\bar{\tau}$. This provides a concrete dynamical interpretation of tunneling as a rare-event diffusion process underlying the coherent oscillations described by quantum mechanics.

The successful application of the SM formalism to the inversion dynamics of the ammonia molecule further illustrates its predictive power. Using the double Rosen–Morse potential and the formula for the stochastic-mechanical tunneling frequency $\nu_{\rm SM}$ in terms of $\bar{\tau}$ (see Eq.~\eqref{eq:nu_SQ}), we obtained an inversion frequency of approximately $24~\mathrm{GHz}$, in remarkable agreement with the experimental value. This quantitative concordance demonstrates that the SM framework captures not only the qualitative features of tunneling dynamics but also its measurable spectroscopic signature. In this sense, the stochastic representation is not merely an interpretational alternative but a dynamically consistent formulation of quantum mechanics at the level of Schrödinger evolution and single-time observables, while offering practical computational advantages in certain contexts.

%

\section{Conclusions}

The present work demonstrates that Nelson’s stochastic mechanics provides a coherent and quantitatively accurate framework for describing quantum tunneling as a diffusive process in real space and time. By establishing an explicit link between the mean stochastic tunneling time and the quantum‐mechanical oscillation period, we have shown that SM can reproduce measurable quantum timescales while offering a more intuitive dynamical picture based on stochastic trajectories. The derived relation $\tau_{\mathrm{QM}} = (\pi/2)\,\bar{\tau}$, validated both analytically and numerically, reveals a universal correspondence between the stochastic and quantum descriptions in the high‐barrier limit.

The stochastic-mechanics formalism has already been extended to several important cases beyond the one-dimensional systems studied here. These include dissipative quantum systems \cite{Marra1987,Bacciagaluppi2012}, multi-particle systems exhibiting quantum nonlocality \cite{delaPena1969,Loffredo2007}, and semiclassical dynamics \cite{JonaLasinio1981}.  A comprehensive modern treatment of stochastic mechanics, including relativistic extensions and applications on curved spacetimes, has been recently provided by Kuipers \cite{Kuipers2023}. By providing direct access to first-passage-time distributions---quantities not easily obtained from the standard quantum formalism---stochastic mechanics may offer complementary insights into tunneling processes in complex molecular and condensed-matter systems.

In summary, our findings strengthen the physical foundations of stochastic mechanics and highlight its ability to yield quantitatively precise and conceptually transparent descriptions of quantum tunneling, bridging the gap between stochastic dynamics and coherent quantum evolution.

\section*{Acknowledgments}  
This work was supported in part by the following Brazilian agencies: Conselho Nacional de Desenvolvimento Cient\'ifico e Tecnol\'ogico (CNPq), under Grant Numbers 307385/2023-0 (GLV) and 307626/2022-9 (AMSM) and Coordena\c c\~ao de Aperfei\c coamento de Pessoal de N\'ivel Superior (CAPES).

\appendix

\section{Unitary Equivalence to Quantum Mechanics}
\label{sec:unitary_equivalence}

While the hydrodynamic equivalence discussed in Sec.~\ref{sec:review} establishes a correspondence between Nelson’s stochastic mechanics framework and quantum mechanics through the Madelung flow equations, further insight can be gained by examining the operator formulation of this equivalence. This alternative perspective reveals that the stochastic and quantum evolutions are actually related through a unitary 
transformation, which provides a more profound understanding of the mathematical structure 
underlying the correspondence between the two formalisms.

The key observation here is that the forward and backward stochastic processes can be 
combined into a single complex-valued drift, leading to what is called a bi-directional generator 
\cite{Pavon2000}. To see how this works, let us first introduce a complex-valued 
\emph{quantum drift velocity} defined by
\begin{align}
\mathbf{v}_q(\mathbf{x}, t) = \mathbf{v}(\mathbf{x}, t) - i\mathbf{u}(\mathbf{x}, t) = 
\frac{\hbar}{mi} \nabla \ln \psi(\mathbf{x}, t),
\label{eq:quantum_drift}
\end{align}
where we have used Eqs.~(\ref{eq:vIm}) and (\ref{eq:uRe}). This complex drift velocity 
encodes all the essential quantum mechanical information in a compact form.

Now, let us define the forward and backward generators, $L_\pm$, acting on smooth, compactly supported 
test functions $f(\mathbf{x}, t)$, as the spatial–derivative components of the total time-derivative operators $D_\pm$ [see Eq.~(\ref{eq:Df})]:
\begin{align}
(L_\pm f)(\mathbf{x}, t) = \mathbf{v}_\pm(\mathbf{x}, t) \cdot \nabla f(\mathbf{x}, t) \pm 
\frac{\sigma^2}{2} \nabla^2 f(\mathbf{x}, t),
\end{align}
where we recall that $\sigma^2 = \hbar/m$. Using the complex drift velocity $\mathbf{v}_q$ defined above, we 
can construct a \emph{bi-directional generator}
\begin{align}
(L_b f)(\mathbf{x}, t) = \mathbf{v}_q(\mathbf{x}, t) \cdot \nabla f(\mathbf{x}, t) -
\frac{i\sigma^2}{2} \nabla^2 f(\mathbf{x}, t),\label{eq:bidirectional_generator}
\end{align}
which can be decomposed as $L_b = \frac{1-i}{2}L_+ + \frac{1+i}{2}L_-$. This operator 
captures the inherently bi-directional  nature (in time) of the stochastic processes.

The central result that establishes the operator equivalence can be stated as follows 
\cite{Pavon2000}. Let $H = -\frac{\hbar^2}{2m}\nabla^2 + V(\mathbf{x})$ be the Hamiltonian 
operator and $\psi(\mathbf{x}, t)$ be a solution of the Schrödinger equation
\begin{align}
i\hbar \frac{\partial \psi}{\partial t} = H\psi.
\label{eq:schrodinger_repeat}
\end{align}
We assume throughout that $\psi$ is never vanishing on the time interval considered and satisfies 
the finite-action condition $\int\!\!\int |\nabla\psi|^2\,dx\,dt < \infty$. Under these 
conditions, one can show \cite{Pavon2000} that the stochastic evolution operator $\partial_t + L_b$ is unitarily 
equivalent to the quantum evolution operator $\partial_t + \frac{i}{\hbar}H$ via the 
conjugation
\begin{align}
\partial_t + L_b = M_\psi^{-1} \left(\partial_t + \frac{i}{\hbar}H\right) M_\psi,
\label{eq:unitary_equivalence}
\end{align}
where $M_\psi$ is the multiplication operator $(M_\psi g)(\mathbf{x},t)=\psi(\mathbf{x},t)\,g(\mathbf{x},t)$. 

It is worth noting that when the wavefunction $\psi$ possesses nodes (zeros), additional care is required, since the osmotic velocity involves gradients of $\log|\psi|$, which become singular on the nodal set. In such cases, the configuration space may be decomposed into nodal domains, and the correspondence between stochastic mechanics and the Schrödinger formulation can be extended through appropriate regularization procedures. For most practical purposes, however---as in the examples considered in the main text---, working with nodeless wavefunctions suffices, in which case the mapping between the two formulations is technically straightforward. In this setting, the operator $M_\psi$ establishes a unitary correspondence between the Hilbert-space representation and its stochastic counterpart.

An important practical observation is that we only need a single reference state to establish 
this connection. To see this, let $\psi$ be a never-vanishing solution of the Schrödinger equation for $H$ 
on the time interval $I$ (for example, a nodeless ground state). Then the intertwining identity
\[
\partial_t+L_b(\psi)=M_\psi^{-1}\Big(\partial_t+\tfrac{i}{\hbar}H\Big)M_\psi
\]
holds and for every Schrödinger solution $\phi$ the ratio
$f=\phi/\psi$ solves $(\partial_t+L_b(\psi))f=0$. In particular, if
$H\phi_n=E_n\phi_n$ are energy eigenstates and $\psi(x,t)=\psi_0(x)e^{-iE_0 t/\hbar}$ is a 
stationary and nodeless state, then
\[
f_n(t,x)=\frac{\phi_n(x)}{\psi_0(x)}\,e^{-i(E_n-E_0)t/\hbar}
\]
solves $(\partial_t+L_b(\psi))f_n=0$. Thus all the spectral information of $H$ can be 
transported to the stochastic representation via $M_\psi$, without requiring a different map for 
each state.
The unitary equivalence has profound physical implications. Any quantum mechanical observable 
$\hat{A}$ corresponds to a stochastic observable $\tilde{A} = M_\psi^{-1} \hat{A} M_\psi$. 
At each fixed time $t$, expectation values are preserved ensuring that all physical predictions remain unchanged between the two formulations.

\bibliographystyle{apsrev4-2}
\bibliography{references}

\end{document}